\documentclass[acmsmall]{acmart}

\usepackage{mdframed}
\usepackage{float}
\usepackage{booktabs}
\usepackage{caption}
\usepackage[fixed]{fontawesome5}
\faStyle{regular}
\usepackage[shortlabels]{enumitem} 
\usepackage{subcaption}
\usepackage{graphicx}
\usepackage{adjustbox}
\usepackage{changepage}
\usepackage{enumerate}  
\usepackage{array, colortbl}

\AtBeginDocument{%
  \providecommand\BibTeX{{%
    \normalfont B\kern-0.5em{\scshape i\kern-0.25em b}\kern-0.8em\TeX}}}

\setcopyright{acmlicensed}
\acmJournal{PACMHCI}
\acmYear{2024} \acmVolume{8} \acmNumber{CSCW2} \acmArticle{499} \acmMonth{11}\acmDOI{10.1145/3687038}

\begin{document}

\title[Guiding Students in Using LLMs]{Impact of Guidance and Interaction Strategies for LLM Use on Learner Performance and Perception}

\author{Harsh Kumar}
\email{harsh@cs.toronto.edu}
\orcid{0000-0003-2878-3986}
\affiliation{%
  \institution{University of Toronto}
  \streetaddress{40 St. George Street}
  \city{Toronto}
  \state{Ontario}
  \country{Canada}
  \postcode{M5S 2E4}
}

\author{Ilya Musabirov}
\email{imusabirov@cs.toronto.edu}
\affiliation{%
  \institution{University of Toronto}
  \country{Canada}
}

\author{Mohi Reza}
\email{mohireza@cs.toronto.edu}
\affiliation{%
  \institution{University of Toronto}
  \country{Canada}
}

\author{Jiakai Shi}
\email{jiakai.shi@mail.utoronto.ca}
\affiliation{%
  \institution{University of Toronto}
  \country{Canada}
}

\author{Xinyuan Wang}
\email{xyben.wang@mail.utoronto.ca}
\affiliation{%
  \institution{University of Toronto}
  \country{Canada}
}

\author{Joseph Jay Williams}
\email{williams@cs.toronto.edu}
\affiliation{%
 \institution{University of Toronto}
  \country{Canada}
  }

\author{Anastasia Kuzminykh}
\email{anastasia.kuzminykh@mail.utoronto.ca}
\affiliation{%
  \institution{University of Toronto}
  \country{Canada}
}

\author{Michael Liut}
\email{michael.liut@utoronto.ca}
\orcid{0000-0003-2965-5302}
\affiliation{%
  \institution{University of Toronto Mississauga}
  \country{Canada}
}
\renewcommand{\shortauthors}{Harsh Kumar, et al.}

\begin{abstract}
Personalized chatbot-based teaching assistants can be crucial in addressing increasing classroom sizes, especially where direct teacher presence is limited. Large language models (LLMs) offer a promising avenue, with increasing research exploring their educational utility. However, the challenge lies not only in establishing the efficacy of LLMs but also in discerning the nuances of interaction between learners and these models, which impact learners' engagement and results. We conducted a formative study in an undergraduate computer science classroom (N=145) and a controlled experiment on Prolific (N=356) to explore the impact of four pedagogically informed guidance strategies on the learners' performance, confidence and trust in LLMs. Direct LLM answers marginally improved performance, while refining student solutions fostered trust. Structured guidance reduced random queries as well as instances of students copy-pasting assignment questions to the LLM. Our work highlights the role that teachers can play in shaping LLM-supported learning environments.


\end{abstract}




\received{January 2024}
\received[revised]{April 2024}
\received[accepted]{May 2024}

\maketitle

\section{Introduction}
\label{section:intro}
The value of personalized attention in enhancing student learning outcomes is well recognized in the contemporary educational landscape \cite{grunwald2015personalized, schwartz2014become, zhang2020understanding, bernacki2021systematic}. For instance, it is well-reflected in the 2 Sigma problem by Benjamin Bloom \cite{bloom19842}, who showed that one-to-one tutored students perform two standard deviations (or ``2 Sigma'') better than students who learned via conventional instructional methods.
However, personalized attention in education is challenged
by the continuous expansion of classrooms, 
and with a growing demand for tailored teaching approaches, there is a notable shortage of human teaching assistants to provide individualized attention \cite{UNESCO2022TeacherShortage, ahmed2023GuardianArticle}, leading to a critical bottleneck in achieving optimal educational impacts.

To address this issue, previous research on Intelligent Tutoring Systems (ITS) and Intelligent Teaching Assistants (ITA) has pioneered adaptive learning experiences and personalized instructional feedback \cite{weitekamp2020interaction, anderson1995cognitive}, increasing the interest of CSCW community to the role that conversational agents can play in supporting the education process \cite{maldonado2023reader, wu2021exploring, breideband2023community, wardrip2013education, liu2020self, sobel2017edufeed, davis2018crowdworkers}.
The rapid emergence of Large Language Models (LLMs) in recent years has created even more promising opportunities for technological support of personalized attention needs of students \cite{zhou2022large, sejnowski2023large}. These models, with their ability to comprehend and generate human-like text, have the potential to reshape the landscape of personalized education. 
Consequently, researchers started exploring the potential advantages of the development of LLM-based agent tutors as potential surrogates for human teaching assistants \cite{rahman2023chatgpt, dergaa2023human, baidoo2023education, kumar2023math}. 
Indeed, the dialogic nature of LLM-based agents, their ability to adapt to individualized learning patterns, and their ease of integration into diverse educational contexts make them a potential boon for educational frameworks \cite{farrokhnia2023swot}. Although learning environments vary between subjects and educational levels, the versatility of LLMs allows their integration into these specialized contexts, providing a unified platform for diverse educational needs. 

However, although there is a considerable volume of existing research on LLM applications, a noticeable gap remains when applied to education. For example, a large body of pedagogical research \cite{keil2000explanation, borko1997teachers, shepard2001role, van2001professional, chi2001learning} suggests that the success of education and tutoring is significantly defined by the communication strategies employed by an instructor. Thus, the pedagogical and interaction effects are inherently interconnected, and discerning their nuanced impacts is essential for the effective design and implementation of LLM tutors.
However, while LLM-based systems can support different interaction strategies \cite{lee2022evaluating, kasneci2023chatgpt}, it remains unclear which of these strategies are most beneficial for an agent to assume when interacting with students. For example, how could these strategies influence the students' reception of the assistance, self-assessment, and their subsequent performance? 
The matter is further complicated by evidence that the dynamics of social interactions might differ for human-human and human-agent communications \cite{doyle2019mapping, clark2019makes}, making the direct integration of the pedagogical knowledge on learning patterns into agent behavior even more challenging. Correspondingly, while there is a substantial body of literature on the impact of human-human teaching interaction strategies on learning outcomes, our understanding of how these dynamics translate to an LLM-learner context remains limited. 
Thus, the question is not only about the feasibility of these LLM-based tutors, but also about the ambiguity surrounding the effects of different interaction strategies on students. To be able to effectively adapt LLM-based agents for individual tutoring, 
we first need to understand what interaction strategies should be implemented in these systems for optimal learning support.

In this work, we address this gap by exploring the following overarching research question:

\noindent\textbf{RQ} \emph{What are the specific effects of 
common guidance strategies (
providing instructions, presenting examples, employing metacognitive questioning, and encouraging initial problem-solving with subsequent refinement) on the learning process supported by an LLM tutor?}

In order to answer this general question, we formulate a set of more specific RQs:
\vspace*{1em}
\begin{description}
\hrule height 0.1em
    \item[RQ1] What differences in learners’ conversation patterns with an LLM are observed under varying guidance given to the learner?
    \item[RQ2] What measurable impact do these guidance strategies have on learners’ performance in learning tasks while using the LLM?
    \item[RQ3] What influence do various guidance strategies have on enhancing or reducing learners' confidence in their ability to use LLMs for problem solving at each stage of interaction?
    \item[RQ4] How do guidance strategies modify learners’ trust in LLM responses through the course of different interaction stages?
    \item[RQ5] In what specific ways do different guidance methods shape learners' self-confidence in problem solving across the various stages of interaction?
\end{description}
\hrule height 0.025em
\vspace*{0.5em}

To gain a deeper understanding of student interactions with LLMs in educational contexts, in this paper, we distinguish three key elements of student-agent interactions: the \textbf{Guidance Received}, which refers to the directions or suggestions students obtain for using LLMs; the \textbf{Learner Approach}, indicating whether learners first consult the LLM or attempt problem solving independently; and the \textbf{LLMs' Response}, emphasizing the importance of dialogue quality with LLM. These components are intertwined, with shifts in one potentially influencing the others. 
We then empirically investigated these influences through two user studies. In both studies, we explored \textit{four} variations of the first element, using four distinct guidance strategies: list-based suggestions, example-based instruction, metacognitive questioning (making learners think about their use of LLMs before actual use) and a problem-solving-then-refinement approach (making learners solve a problem first and then refine it with LLMs). 
These four guidance strategies manifested in \textit{two} distinct student approaches (second element): directly consulting the LLM or relying on self-first problem solving. 
Finally, we manipulated the third element -- LLMs' Response -- through pre-interaction manipulations \cite{brown2020language, kumar2022exploring}. 




Although there is previous research on student-AI interaction, LLMs seem to provide different levels of interaction experience and convincingness, even in the presence of errors. Thus, studying the multidimensional picture of student perceptions and their dynamics and analyzing them in a formative experimental setup, combining preliminary quantitative measures as well as extensive qualitative feedback is needed at this initial stage. Our formative study in an undergraduate computer science classroom ($N=145$) tasked students with solving an assignment while receiving help from an LLM-based chatbot. We explored the impact of our \textit{four} guidance strategies, and the resulting \textit{two} approaches, on the student's performance (in the assignment, as well as in the final exam 2 weeks after the assignment), system perception (helpfulness, confidence in LLM's response, tolerance of mistakes, willingness to interact again), and their self-assessment (change in their self-efficacy, confidence in their answers). We found a reduction in unrelated student queries to LLM with structured guidance such as \textit{List of Suggestions}, \textit{Metacognitive Questioning}, and \textit{Solve then refine with LLM} approaches, while \textit{Example based instruction} paradoxically increased such queries. Furthermore, certain types of guidance promoted deeper engagement with the material, evidenced by fewer instances of verbatim question copying and increased rephrasing, highlighting the impact of guidance on the quality and focus of student-tutor interactions in LLM environments. As we discuss further, from a statistical point of view, this setting is traditional in its inherent limitations of sample size by course enrollment, student dropout, and selective engagement. It allows us to use it only to gather formative data for different conditions. In addition, exposing students to experimental stimuli allows us to gather qualitative feedback specific to different stimuli.

Informed by findings from formative deployment, we conducted a controlled study on Prolific \cite{palan2018prolific} in which crowdworkers (N=356) were asked to solve math problems at the elementary and high school level with the help of an LLM chatbot. This math knowledge level was selected to ensure a standardized baseline of complexity and familiarity for a broad adult audience, allowing for a more controlled and consistent assessment of the LLM chatbot's tutoring capabilities. Besides manipulating the different forms of guidance given to students, the online controlled setting allowed us to compare a pre-prompted LLM with an unprompted LLM. Through a 2 (\textit{LLM Chatbot}: Prompted vs. Unprompted) x 5 (\textit{Guidance strategy}: G1-List of Suggestions vs. G2-Example-based Instruction vs. G3-Metacognitive Questioning vs. G4-Solve then refine with LLM vs. No Instruction), we explored the impact of interaction strategies on the learners' performance and their confidence (in LLM's responses, ability to use LLMs, and their Math-solving ability) across the different stages of interaction (before receiving guidance, immediately after receiving guidance, and after task completion). We did not find statistically significant differences in Learner + AI task performance across different Guidance Types and LLM Types (prompted vs. unprompted), and we observe interesting patterns in confidence in the ability to use LLMs to solve math problems and trust in LLM responses over different stages of interaction. The guidance briefly boosts confidence in the ability to use LLM, but both measures significantly drop after learners attempt to solve real tasks using LLM. This might suggest a partially negative interaction experience in solving tasks, as well as the need for additional learning support for LLM use in the particular context.

Our work informs the design of 
learner-LLM interactions for collaborative learning environments, ensuring the effective and empathetic use of LLMs in pedagogical contexts. 
In the remainder of this paper, we start (Section \ref{section:background}) by describing the related work done in incorporating LLMs in classroom, and highlight the need to explore different LLM interaction strategies. In Section \ref{section:design}, we describe the design space of learner-LLM interactions, and propose interaction designs that are evaluated in Sections \ref{section:field} and \ref{section:mturk}. We conclude by discussing our key findings in the context of existing literature and operationalizing them in a set of design considerations. 

\section{Related Work}
\label{section:background}

In this section, we draw from recent HCI and CSCW research on leveraging conversational agents in education, as well as literature from educational psychology on guidance strategies and self-efficacy to highlight a pressing need to study how learners interact with LLM-based conversational agents in real-world settings and how to guide them using pedagogically informed strategies.

\subsection{Understanding Learner-LLM Interactions in Real-World Settings}

With LLM-based conversational systems like ChatGPT \cite{chatgpt}, Bard \cite{bard}, and BingChat \cite{bingchat} already out in the wild and used daily by hundreds of thousands of learners, educators and researchers face an urgent need to understand how learners interact with these systems in real-world educational settings. They are both thrilled \cite{rahman2023chatgpt, dergaa2023human, baidoo2023education} and concerned \cite{rahman2023chatgpt, dergaa2023human} about the integration of LLM-based chatbots into diverse learning environments as we navigate the uncertainties regarding the impact of this rapidly evolving technology on pedagogy. 

The growing body of recent CSCW and HCI literature on the incorporation of LLM-based conversational agents in education has explored various ways to enhance chatbots to assist learners in different ways, such as using them as persuasive agents to target behavior change through self-reflection \cite{li2023exploring}, making them more goal-aware \cite{deng2023rethinking}, proactive \cite{liao2023proactive}, and credible \cite{candello2023means}, and customizing them to take on different roles to facilitate different discussion patterns and system thinking \cite{nguyen2023role}. However, while such works can provide valuable preliminary insights into how we can design chatbots, they often do not investigate learner interactions in real-world settings. Given that learners are already interacting with widely available and powerful LLMs through tools like ChatGPT, it has become increasingly important to understand how learners are interacting with these tools. In this work, we offer empirical insights of classroom and crowdworker interactions with LLMs from two field studies that contribute to this understanding, and offer design recommendations that can inform the design of future LLM-based chatbot implementations.

\subsection{Leveraging Guidance Strategies to Steer Interactions Toward Learning}
\label{section:guidance_lit}

In conjunction with investigating learner-LLM interactions, we also need to understand how to guide students in ways that support their learning rather than impede it. Guidance plays a crucial role in helping users navigate how to best interact with chatbots, but the efficacy of specific forms of guidance remains underexplored \cite{wu2021exploring}. 

The CSCW community has begun to explore different aspects of this important space. For example, Wu et al. \cite{wu2021exploring} studied user preferences for two different types of guidance (Example-Based and Rule-Based) at four different timings of guidance (Service-Onboarding, Task-Intro, After Failure and Upon Request) based on data from 24 participants, and found that users preferred Example-Based guidance, and wanted to see the guidance as part of Task-Intro. Zhu et al. \cite{zhu2022action} built a chatbot prototype that provided task-based instructions to novice workers, and offered results from a pilot study with 7 participants that indicated that back-and-forth conversations could help novice workers follow instructions, and perceive those instructions as more actionable. Our work contributes to this body of literature on how different guidance strategies impact learner perceptions and interactions with LLM-based chatbots. A distinguishing feature of our work from existing preliminary studies is that we deployed our experiments in two distinct learner groups with comparatively larger sample sizes (145 students in an undergraduate CS classroom and 356 crowdworkers from Prolific). We study the specific effects of four guidance strategies (providing instructions, presenting examples, using metacognitive questioning, and encouraging initial problem solving with subsequent refinement).

To inform our selection of guidance strategies, in addition to looking at the types of guidance that have been explored in preliminary work within the CSCW community, we also consulted previous pedagogical literature on various forms of instructional guidance \cite{bonawitz2011double}, and chose to study direct instruction \cite{white1988meta}, example-based instruction \cite{wittwer2010effective}, and metacognitive strategies \cite{livingston2003metacognition, csen2009relationsip}—such as posing reflective questions \cite{ellis2014analysis, alt2020reflective}, due to their documented effectiveness in prior research \cite{gersten1986direct, adams1996research, zendler2018effect, anderson2007supporting, van2010example, palincsar1986metacognitive, bourner2003assessing}. 

To understand why guidance is of particular relevance to LLM-based chatbot design, we turn to recent research on chatbot interactions of novice AI users that show prompt-writing can be deceptively difficult. The response from LLM-based chatbots is primarily steered by text-based prompts \cite{wu2022ai, zhou2022large}. However, recent research has indicated that crafting effective prompts is challenging \cite{zamfirescu2023johnny}, particularly for those without a deep understanding of AI \cite{abdellatif2020challenges, zamfirescu2023johnny, fiannaca2023programming}. The conversational interface of these chatbots mimics human interaction \cite{smestad2019chatbot}, potentially misleading students into thinking that prompt-writing is as straightforward as talking to humans \cite{hofstadter1995fluid}. This gap between perceived simplicity and the actual complexity of effective prompt-writing can breed overconfidence. However, when the chatbot fails to respond as expected, students' trust in the AI system \cite{denny2023can} can wane, leading to frustration \cite{perez2020rediscovering}. In response to these challenges, informal communities like the r/aipromptprogramming and r/ChatGPT subreddits have emerged to share best practices for prompt-writing. Simultaneously, researchers are starting to establish formal guidelines \cite{denny2023promptly, liu2022design, ekin2023prompt}, but given the nascent nature of this field, much remains to explore. To improve student success with LLM, it is necessary to convert these guidelines into a more digestible format, thereby improving LLM literacy \cite{zamfirescu2023johnny, yang2023use}. 

This paper investigates the creation of these student-centric guidelines, translating them into clear instructions for LLM use. More specifically, we evaluate how various guidance strategies—including direct instruction, example-based teaching, metacognitive questioning, and worked examples—affect students' performance, self-confidence, trust in the chatbot, and their perceptions of its efficacy.

\subsection{Understanding Factors that Influence Learners’ Help-Seeking Decisions}
\label{section:behavior_lit}
To characterize how learner-LLM interactions may be shaped by important factors that influence learner help-seeking decisions more broadly, we can turn to the educational psychology literature on self-efficacy and perceived task difficulty. The choice of when to consult external tools during problem-solving isn't exclusive to the realm of LLMs. In broader educational research, students' decisions to access resources—whether they are textbooks, online forums, peers, or tutors—often come down to an interplay between self-efficacy, perceived task difficulty, and the accessibility of the tool.

\textit{Self-Efficacy and Task Difficulty.} Research has shown that learners with high self-efficacy and (over-) confidence can influence learning behavior \cite{moores_self-efficacy_2009,vancouver_two_2002}. High self-efficacy students might first attempt to tackle problems independently. However, if they perceive the task as excessively challenging or outside of their competence, they are more likely to consult external sources earlier in their problem-solving process.

\textit{Availability and Perceived Utility of the Tool.} The ease of accessing a tool and the learner's belief in its utility can impact when and how often they use it. For instance, the popularity of online forums in CS courses can be attributed to the instantaneous, community-validated feedback they provide \cite{glassman2015overcode}. 

\textit{Balancing Independence and Reliance.} The idea of balancing self-reliance with tool reliance fits well in the framework of self-regulated learning \cite{karabenick_understanding_2011}. In many settings, students are trained to strike a balance, ensuring that they understand the basic concepts, while not hesitant to seek help when needed.

The aforementioned considerations have nuanced significance in the context of LLMs. The availability and perceived efficacy of LLMs could encourage more immediate reliance, potentially overshadowing traditional problem-solving approaches. Our paper's exploration of different student approaches offers insight into these nuanced behaviors in the specific context of LLM interactions.

We factored such considerations into the design of our experiments. We included questions that help us quantify such factors using pre- and post-measures regarding perception of LLMs (confidence in LLMs' responses, helpfulness, willingness to interact again, and error tolerance) and self-perception (student confidence in their answers, self-confidence for the given topic of the assignment). To measure long-term learning, we assessed their performance on the final exam on the same topic, comprising isomorphic questions to those in the assignment.

\section{Design Considerations for Learner Interaction with LLMs}
\label{section:design}

Understanding how students interact with LLM-based tools becomes imperative as language models continue to find their way into educational settings. In our exploration of learner interactions with LLMs, we identified three critical factors influencing these engagements:

\paragraph{1. Guidance Received} The directions or suggestions that students receive can shape their entire interaction experience. The form and content of this guidance can determine how students approach and utilize LLMs to their advantage (Section \ref{section:guidance_types}). 

\textbf{Motivation:} Recent research highlights the challenges in prompting LLMs for specific tasks \cite{zamfirescu2023johnny}, emphasizing the need for structured guidance to support effective use \cite{abdellatif2020challenges, fiannaca2023programming}. This is supported by extensive literature that indicates that appropriate guidance can help learners form the correct mental models of tools \cite{kieras1984role}, thus facilitating their proper use. Despite the critical importance of guiding learners in using LLMs, this remains a largely unexplored problem (see Section \ref{section:guidance_lit}). We address this gap by exploring guidance strategies for LLM use.

\paragraph{2. Learner Approach} Learners' initial inclination to consult the LLM immediately or first wrestle with the problem alone can have implications for their learning process. The strategies they adopt can be influenced by the guidance they have received and can vary significantly in efficacy (Section \ref{section:use_types}).

\textbf{Motivation:} Existing research indicates that individual factors such as self-efficacy, prior exposure, and perceived utility can influence users' approach when using tools \cite{moores_self-efficacy_2009,vancouver_two_2002, glassman2015overcode}. Investigating these approaches in the context of LLM tutors is important, as forming the appropriate reliance and dependency on these tools is necessary for effective learning \cite{karabenick_understanding_2011} (see Section \ref{section:behavior_lit} for details). Understanding the approaches students take when using LLMs and how guidance can steer these approaches is vital. Some approaches may not be conducive to learning, such as using LLMs to cheat versus using them to seek feedback \cite{kumar2023math}. Hence, we consider the learner's approach a crucial design consideration in our study.

\paragraph{3. LLMs' Response} While guidance and approach set the stage, the actual dialogue with the LLM remains paramount. The quality, clarity, and relevance of the LLMs' responses can make or break the learning experience (Section \ref{section:steer_llm}).

\textbf{Motivation:} Research suggests that the accuracy and framing of responses by conversational agents can significantly influence user perception and interaction with the agent \cite{diederich2022design, xiao2007role}. This is particularly crucial in educational settings where how answers are framed can impact long-term learning outcomes \cite{weber2021pedagogical}, even if the immediate effect seems beneficial (e.g., students using LLMs for cheating) \cite{kumar2023math}. By exploring the steering of LLM responses, we can gain insights into the differences between publicly available LLMs, such as ChatGPT, and those configured by instructors. Additionally, the nature of LLM responses can influence learners' trust in the LLM and their sense of self-efficacy \cite{park2023utilizing, riefle2022may}, as direct answers might lead to a realization of lacking knowledge \cite{yildiz2023conversational}. These factors prompted us to consider the types of LLM responses as a design element in this work.

Each of these components does not exist in isolation but intertwines and influences the others. It is a dynamic relationship where modifying one aspect might lead to shifts in the others. Grasping these intricate relationships is key for educators, designers, and policy makers who want to harness the full potential of LLMs in the learning environment.

\subsection{Guidance Strategies for LLM Engagement}
\label{section:guidance_types}
\begin{figure}
  \includegraphics[width=\textwidth]{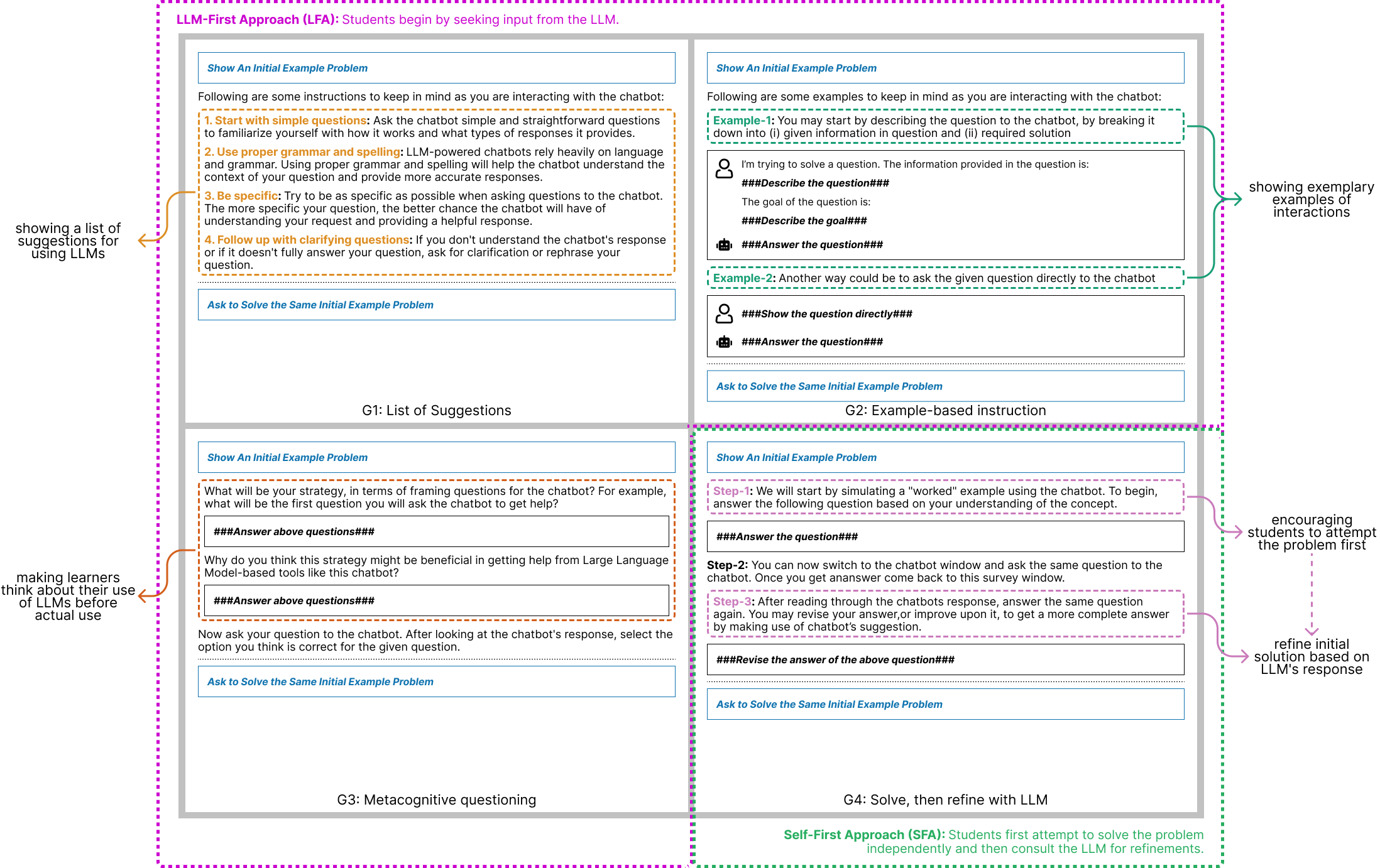}  
  \caption{Different forms of guidance strategies.}  
  \label{fig:guidance}
\end{figure}

In Section \ref{section:guidance_lit}, we discussed how our guidance strategies drew insights from previous work in the pedagogical literature \cite{bonawitz2011double, gersten1986direct, adams1996research, zendler2018effect, anderson2007supporting, van2010example, palincsar1986metacognitive, bourner2003assessing}. In this section, we revisit some of those insights and describe the specifics of our design and the rationale behind each guidance strategy. Figure \ref{fig:guidance} shows the format of each guidance strategy used in our study.
\vspace*{1em}
\begin{description}
\hrule height 0.1em
\item[G1: List of Suggestions] We devised a set of suggestions for the use of LLM chatbots, as depicted in Figure \ref{fig:guidance}A. Given the limited existing research on LLM guidance for students, our suggestions were based on prominent online sources \cite{Pangu_2023, Miquelino, Mollick_2023e}. Recognizing the growing classroom trend of prescribed LLM instructions \cite{ethan_2023, Mollick_2023a}, G1 serves as a representative model to evaluate its efficacy against other guidance strategies.

\item[G2: Example-based instruction] Leveraging examples is an intuitive approach to tool introduction \cite{lee2004work}. However, the choice of examples plays a crucial role. For our classroom deployment, we took examples of interactions from previous interview studies with students \cite{kumar2023quickta, xiaopreliminary}. These examples, validated by an instructor and a teaching assistant, showcased interactions where students found the LLM beneficial. Participants were presented with two such exemplary interactions, as illustrated in Figure \ref{fig:guidance}B.

\item[G3: Metacognitive questioning] (making learners think about their use of LLMs before actual use). Framing the initial question effectively is crucial when seeking help, more so with LLM chat tools where the initial query shapes subsequent responses \cite{daniel2017thinking, zamfirescu2023johnny}. To enable this skill, participants were presented with an example problem and prompted to draft their opening question to the LLM chatbot. Subsequently, they reflected on the potential benefits of their chosen approach (Figure \ref{fig:guidance}C).

\item[G4: Solve then refine with LLM] Worked examples, step-by-step illustrations of the process required to complete a task, are established effective learning tools \cite{sweller2006worked}. In this multistep strategy, we first ask the students to solve a problem and provide the solution. They then engaged with the LLM chatbot to refine their initial answers (Figure \ref{fig:guidance}D). 

\end{description}
\hrule height 0.025em
\vspace*{0.5em}

\subsection{Approaches to Leverage LLMs in Problem Solving}
\label{section:use_types}

The manner in which students interact with LLM-based tools is often influenced by the guidance or instructions they receive. The sequence—whether students consult the LLM first or try to solve problems on their own—has implications for both their learning process and the effectiveness of the LLM as a tool. This builds on research in "when" is it better to provide hints for learning \cite{razzaq2010hints}.

\textbf{LLM-First Approach (LFA):} Students begin by seeking input from the LLM. This approach can assist those who need a foundation or direction before delving into problem-solving, ensuring they are on the right track from the onset. However, it may also promote dependency on the LLM, potentially limiting independent problem-solving capabilities.

\textbf{Self-First Approach (SFA):} Students first attempt to solve the problem independently and then consult the LLM for refinements. This method promotes independent thinking and can boost confidence in one's problem-solving abilities. The LLM acts as a secondary check, reinforcing correct methodologies or suggesting improvements. Conversely, there might be a risk that students get stuck or frustrated if their initial attempts are misdirected.

Guidance strategies, as detailed in Section \ref{section:guidance_types}, influence these approaches. For example, G4 explicitly steers students towards the SFA, while G1-3 offer more flexibility in how students choose to engage with the LLM.

\subsection{Response from Large Language Models}
\label{section:steer_llm}
The response generated from proprietary models is perhaps the most difficult to control in our study, due to ever-evolving (behind-the-scene) updates to the models \cite{chen2023chatgpt}. However, the response of a \textit{\textbf{Prompted}} LLM definitely differs from an \textbf{\textit{Unprompted}} LLM. A few-shot-prompted LLM can achieve better performance in the task at hand (helping students solve problems) than an uninitiated one \cite{brown2020language}.

\begin{figure}
  \includegraphics[width=\textwidth]{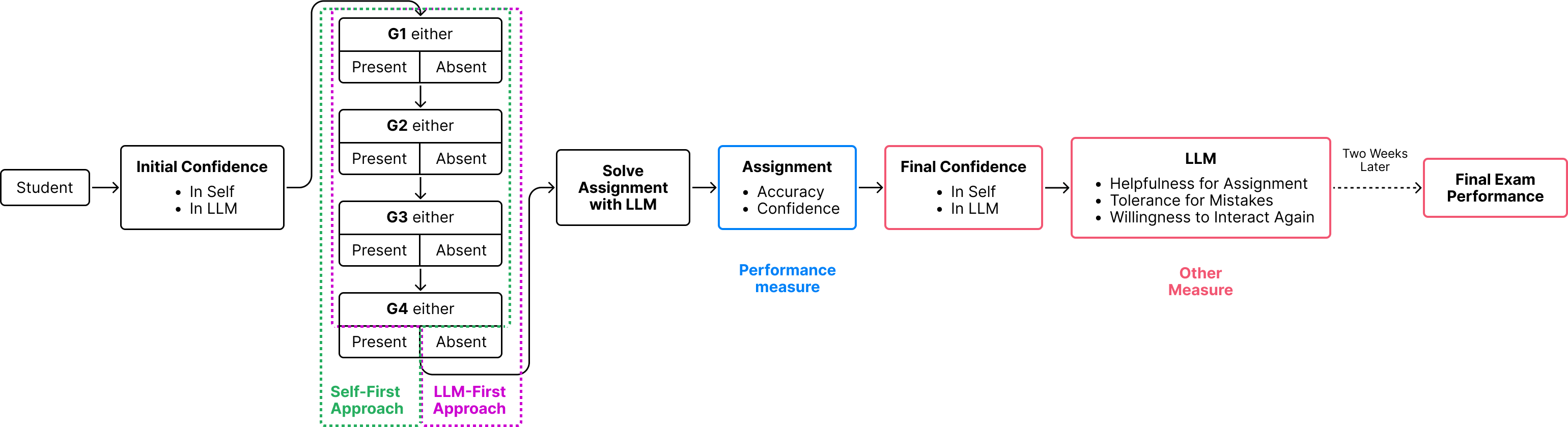}  
  \caption{Schematic of the formative study in a CS classroom. 2 (\textit{List of Suggestions}: present vs. absent) x 2 (\textit{Example-based Instruction}: present vs. absent) x 2 (\textit{Metacognitive-questioning based Instruction}: present vs. absent) x 2 (\textit{Solve, then refine with LLM}: present vs. absent) between subjects.}
  \label{fig:classroom_design}
\end{figure}

To explore this design space, we conducted two studies, one in an upper-year computer science classroom teaching students database-related concepts and a second in an online controlled setting with crowdworkers solving mathematics problems. Figures \ref{fig:classroom_design} and \ref{fig:prolific_design} show a high-level overview of both study designs. They are described in detail in the following sections.

\section{Study-1: A formative field study in a CS classroom}
\label{section:field}

This research study was approved by the local ethics board. The data used in this deployment was obtained from a third-year \textit{Introduction to Databases} course (DB) at a large research-focused North American university in Winter 2023 for Computer Science (CS) students. This course spanned 12 weeks, employing a flipped classroom setup (students were required to complete preparation work: video modules, technical questions, and self-reflections, prior to attending lectures in person using a learning management system). DB had one course instructor, eleven teaching assistants, and was an upper-year elective course for Computer Science students. The majority of students were Computer Science majors, however, some are Computer Science minors. Although natural classroom limitations impose constraints on achievable statistical power and, as a result, the complexity of questions we can reliably answer with quantitative methods, we used this deployment as a formative one, focusing on both acquiring statistical evidence where it is reliable and exploratory analysis without statistical conclusions to inform larger-scale deployments (both in the virtual lab (Study 2, see Section~\ref{section:mturk}) and in the field (future studies)).

We designed a randomized factorial experiment, for the guidance given to the student, with 2 (\textit{G1-List of Suggestions}: present vs. absent) x 2 (\textit{G2-Example-based instruction}: present vs. absent) x 2 (\textit{G3-Metacognitive questioning}: present vs. absent) x 2 (\textit{G4-Solve then refine}: present vs. absent) between subjects. The presence of G4 encouraged a Self-First Approach (SFA), while the absence provided more flexibility (LLM-First Approach -- LFA) (see Section \ref{section:design}). At the beginning of the assignment, students were primed to use LLM-based support tool with the following: 
\begin{quote}
   \textit{``You will now get some instructions related to the use and access of the dialogue-based support system. The system uses a fine-tuned large language model. Large language models (LLMs) are artificial intelligence systems that use deep learning to analyze vast amounts of text and generate responses that are contextually relevant.''}
\end{quote}
After seeing this text, students were randomly assigned to one of the forms of instruction composed of factors described in Section \ref{section:design}. After receiving guidance and access to the LLM chatbot, students attempted to solve multiple-choice problems (related to databases) while receiving support from LLMs. Besides accuracy in the assignment, we collected pre- and post-measures regarding perception of LLMs (confidence in LLMs' responses, helpfulness, willingness to interact again and error tolerance) and perception of self (confidence in their answers, self-confidence for the given topic of assignment). To measure long-term learning, we measured their performance on the final exam on the same topic, comprising isomorphic questions to assignment problems.

\subsection{LLM-based chat support tool}
Each student was given access to a chatbot tutor, based on GPT-3 (described in Section \ref{section:llm_spec}), to help them solve assignment problems. The model prompt was designed after careful iterations between the researchers, the instructor, and the teaching assistant. The prompt given to the model was the following:

\noindent
\begin{mdframed}[backgroundcolor=gray!20]
"The following is a conversation with a database instructor. The instructor helps the human solve assignment problems related to database. The instructor never explicitly gives the solution. The instructor also never writes the SQL query. The instructor would only provide brainstorms to possible solutions without providing any SQL statements. This rule should be enforced in the entirety of the conversation. Following are the DDL files to create table given as part of the assignment."
\end{mdframed}

\subsection{Participants}
There were 218 students enrolled in DB, of which a total of 145 students (67\%) completed the assignment. When students were asked about their familiarity with Large Language Models (LLMs), 51.03\% reported occasional use, 16.55\% were regular users, 15.86\% had tried them once, and 16.55\% had never used them before. On a scale of 1 to 7, the students' average initial self-efficacy for the topic was 4.10 (SD = 1.38).

\begin{table}[ht]
\centering
\caption{Categories of first query asked to the LLM Tutor. The assignment was on the topic of serializability in databases. The first question in the assignment began with \textit{"For each of the following locking protocols (a to d below)..."} and asked students to select the properties that were ensured for the given protocols. Majority of students asked for clarifying specific doubts related to the assignment, while many students also copy-pasted the given assignment problem.}
\resizebox{\textwidth}{!}{\begin{tabular}{p{3cm} p{7cm} p{7cm} p{1cm} }
\toprule
\textbf{Category} & \textbf{Description} & \textbf{Example from Dataset} & \textbf{\% of Students} \\
\midrule
Verbatim First Assignment Question & The student copies the exact question from the assignment verbatim as the first message to the LLM tutor. & \textit{"For each of the following locking protocols (a to d below), assuming that every transaction follows that specific locking protocol..."} & 28\% \\
\midrule
First Assignment Question Reworked & The student rephrases or alters the wording of the first question from the assignment, maintaining its original intent. & \textit{"Given the following, always obtain an exclusive lock before writing; hold exclusive locks until the end of transaction. No shared locks are ever obtained. Which properties are ensured?"} & 14\% \\
\midrule
Unrelated Initial Inquiry/Chatbot Testing & The student's first query is unrelated to the assignment, often to test the chatbot's capabilities or driven by general inquisitiveness. & \textit{"Hi buddy, recite a Shakespearean sonnet..."} & 4\% \\
\midrule
Initial Conceptual Clarification Request & The student seeks to understand specific concepts or parts of the assignment more clearly with their initial query. & \textit{"How do I read the Venn diagram for schedules?"} & 54\% \\
\bottomrule
\end{tabular}}
\label{table:form_qualt}
\end{table}

\subsection{Results}
\subsubsection{Changes in interaction patterns based on guidance type.} 
We analyzed the first query each student sent to the LLM tutor. The goal was to understand how students used LLMs to solve problems and how the guidance received affected the way students used the LLM (RQ1). Of the 145 students who completed the assignment, 138 students used LLMs. We followed an iterative coding scheme to categorize each of the first queries from the students \cite{cooper2012apa}. Two members of the research team went through the entire dataset and after multiple rounds of discussions, arrived at \textit{four} distinct categories as shown in Table \ref{table:form_qualt}. After this, the first coder categorized the entire dataset. The second coder independently labeled a random \( (n = 70) \) subsample., resulting in high inter-rater agreement \( (\text{Cohen's Kappa} = 0.814, Z = 10.7, p < 0.001) \).

\begin{figure}
  \includegraphics[width=\textwidth]{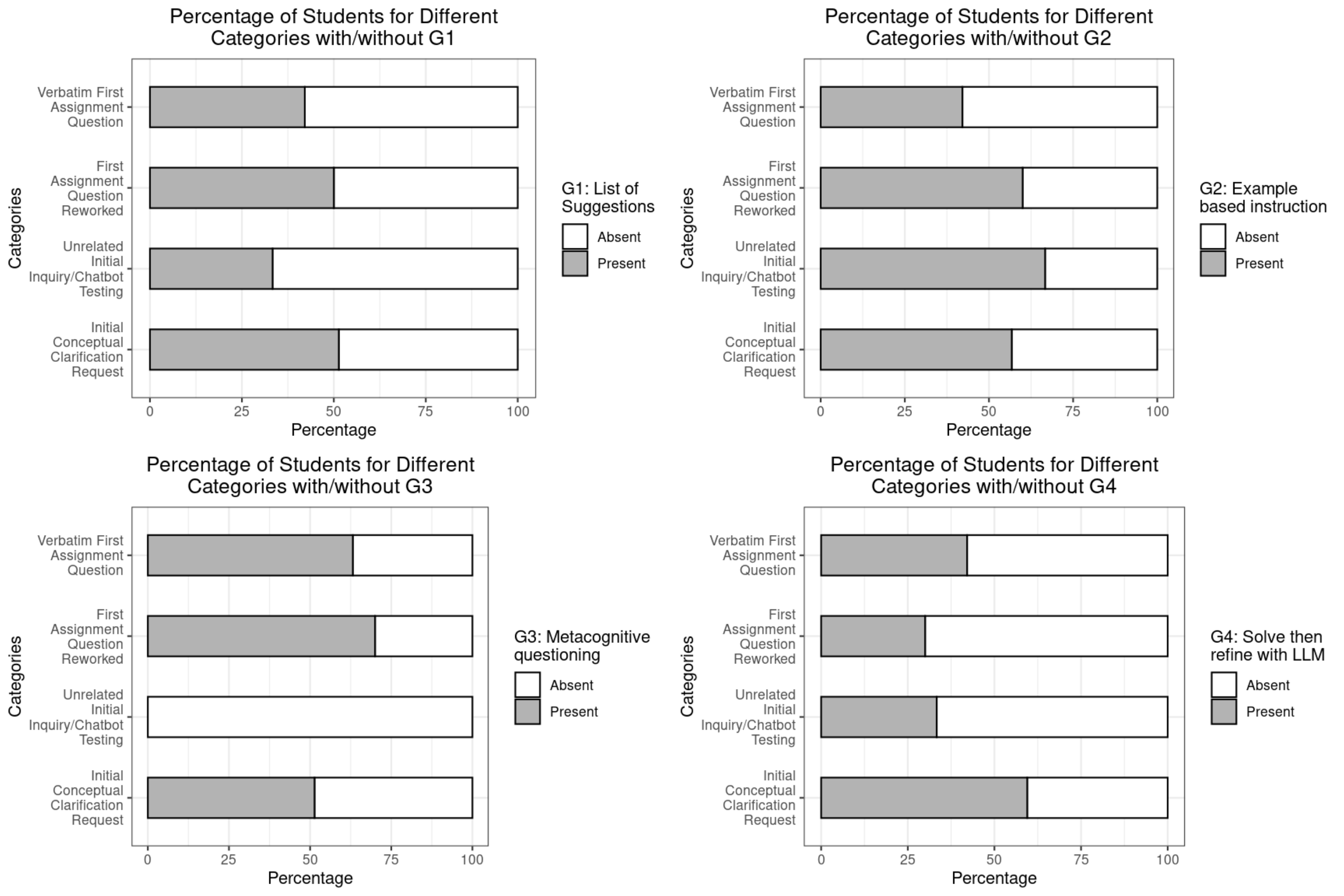}  
  \caption{Distribution of First Query Categories Based on Guidance Exposure. The bars represent the percentage of students whose first query falls into each category, segmented by whether they received a specific type of guidance (G1-G4). The percentages are calculated relative to the total number of students in each category. This visualization helps identify the influence of each guidance type on the students' initial interaction with the LLM. A percentage around 50\% indicates that the presence or absence of the guidance type had little to no discernible impact on the likelihood of a student's first query falling into that category.}
  \label{fig:qual_dist}
\end{figure}

We examined the impact of each guidance type (G1--G4) on the nature of the first queries made by students to the LLM tutor. For each guidance type, students were grouped into two categories: those who received the guidance and those who did not. We then analyzed the first queries made by the students in each group to see how often they fell into each of the established query categories (Figure \ref{fig:qual_dist}). This allowed us to compare the tendencies of the students' initial queries on the basis of whether they had been exposed to a particular type of guidance. We find that when \textit{G1-List of Suggestions}, \textit{G3-Metacognitive Questioning} and \textit{G4-Solve then refine with LLM} are absent, students tend to ask unrelated questions more, compared to when it is present. On the other hand, the presence of \textit{G2-Example based instruction} resulted in more students asking unrelated questions, or trying to break the chatbot. G1, G2, and G4 show a positive effect in reducing the likelihood of the student copy-pasting the exact assignment question. However, G3 showed the opposite effect. The introduction of \textit{G3-Metacognitive Questioning} resulted in a higher percentage of students rephrasing the first assignment question, while the absence of G4 had the same effect. For \textit{G4-Solve then refine with LLM}, we find that students who received this particular guidance asked clarification questions more as their first query, in line with what was instructed to them as part of this strategy.

By aligning the types of guidance with the classified first queries, we were able to identify trends in student interactions. This analysis not only contributed to a deeper understanding of the immediate impact of guidance on student queries, but also offered insights into the broader implications of guidance strategies in shaping student learning experiences with LLMs.

\subsubsection{Students' attitudes towards LLMs.} Students rated their average initial trust in LLMs at 3.58 out of 7 (SD = 1.54). Previous experience of LLM use is connected with the initial trust in LLM ($\chi^2 = 18.186$, $df = 6$, $p < 0.01$), with students having regular experience with LLMs reporting more trust in them. The overall difference in trust towards LLMs (before the experiment) and on LLM chatbots (after the experiment) across balanced experimental conditions shows a significant growth trend (from $\text{Median}_{before} = 4$ to $\text{Median}_{after}=5$ on 7-point Likert scale, \emph{V}(Wilcoxon) = 1421.00, $p < 0.001$)). This gives us insight into the learners' trust in LLMs (RQ4).

\begin{table}
\scriptsize
\setlength{\tabcolsep}{5pt}
\resizebox{\textwidth}{!}{
\renewcommand{\arraystretch}{1.3}
\begin{tabular}{|c|c|c|c|}
\hline
\multicolumn{1}{|c|}{Category} &
  \multicolumn{1}{c|}{Measure} &
  \multicolumn{1}{c|}{Code} &
  \multicolumn{1}{c|}{Range} \\ \hline
Performance &
  Homework Score &
  HS &
  0 - 100 \\ \cline{2-4}
 &
  Exam Score &
  ES &
  0 - 100 \\ \cline{2-4}
 &
  Score Diff &
  SD &
  (-100) - 100 \\ \hline
  Perception of System &
  Helpfulness &
  H &
  1 - 7 \\ \cline{2-4}
 &
  Helpfulness for Other Topics &
  HOT &
  1 - 7 \\ \cline{2-4}
 &
  Tolerance for Mistakes &
  TM &
  1 - 7 \\ \cline{2-4}
 &
  Willingness to Interact Again &
  WIA &
  1 - 7 \\ \cline{2-4}
 &
  \begin{tabular}[c]{@{}l@{}}Change in Confidence in\\ the LLM's Ability to Help\end{tabular} &
  CCLAH &
  (-14) - 14 \\ \hline  
  Perception of Self &
  Change in Self-confidence &
  CS &
  (-14) - 14 \\ \cline{2-4}
 &
  Confidence on Their Answers &
  CTA &
  1 - 7 \\ \hline
\end{tabular}}
\caption{List of outcome measures with their scales in Study-1. We assign a code to each measure so that they are easy to reference at different points in the paper.}
\label{table:study1-measures}
\end{table}

\begin{table}
\scriptsize
\setlength{\tabcolsep}{5pt}
\resizebox{\textwidth}{!}{
\renewcommand{\arraystretch}{1.3}
\begin{tabular}{|c|cc|cc|cc|cc|}
\hline
\multicolumn{1}{|c|}{Code} &
  \multicolumn{2}{c|}{G1} &
  \multicolumn{2}{c|}{G2} &
  \multicolumn{2}{c|}{G3} &
  \multicolumn{2}{c|}{G4} \\ \hline
   &
  \multicolumn{1}{c|}{Absent} &
  \multicolumn{1}{c|}{Present} &
  \multicolumn{1}{c|}{Absent} &
  \multicolumn{1}{c|}{Present} &
  \multicolumn{1}{c|}{Absent} &
  \multicolumn{1}{c|}{Present} &
  \multicolumn{1}{c|}{Absent} &
  \multicolumn{1}{c|}{Present} \\ \hline
  HS &
\multicolumn{1}{l|}{$58.69(\pm 2.09)$} &
\multicolumn{1}{l|}{\cellcolor{green!25}$59.46(\pm 1.85)$} &
\multicolumn{1}{l|}{$57.52(\pm 1.81)$} &
\multicolumn{1}{l|}{\cellcolor{green!25}$60.62(\pm 2.10)$} &
\multicolumn{1}{l|}{$57.63(\pm 1.87)$} &
\multicolumn{1}{l|}{\cellcolor{green!25}$60.47(\pm 2.05)$} &
\multicolumn{1}{l|}{$61.38(\pm 1.83)$} &
\multicolumn{1}{l|}{\cellcolor{red!25}$56.87(\pm 2.06)$}
   \\ \hline
  ES &
\multicolumn{1}{l|}{$56.20(\pm 1.52)$} &
\multicolumn{1}{l|}{$54.80(\pm 1.70)$} &
\multicolumn{1}{l|}{$53.61(\pm 1.74)$} &
\multicolumn{1}{l|}{$57.33(\pm 1.45)$} &
\multicolumn{1}{l|}{$55.49(\pm 1.79)$} &
\multicolumn{1}{l|}{$55.47(\pm 1.44)$} &
\multicolumn{1}{l|}{$55.00(\pm 1.43)$} &
\multicolumn{1}{l|}{$55.95(\pm 1.78)$}
   \\ \hline
  SD &
\multicolumn{1}{l|}{$-2.49(\pm 2.34)$} &
\multicolumn{1}{l|}{$-4.66(\pm 2.23)$} &
\multicolumn{1}{l|}{$-3.91(\pm 2.14)$} &
\multicolumn{1}{l|}{$-3.29(\pm 2.42)$} &
\multicolumn{1}{l|}{$-2.14(\pm 2.19)$} &
\multicolumn{1}{l|}{$-5.00(\pm 2.37)$} &
\multicolumn{1}{l|}{$-6.38(\pm 1.97)$} &
\multicolumn{1}{l|}{$-0.92(\pm 2.51)$}
   \\ \hline
  H &
\multicolumn{1}{l|}{$4.51(\pm 0.22)$} &
\multicolumn{1}{l|}{$4.82(\pm 0.16)$} &
\multicolumn{1}{l|}{$4.56(\pm 0.19)$} &
\multicolumn{1}{l|}{$4.78(\pm 0.19)$} &
\multicolumn{1}{l|}{$4.70(\pm 0.19)$} &
\multicolumn{1}{l|}{$4.64(\pm 0.20)$} &
\multicolumn{1}{l|}{$4.54(\pm 0.19)$} &
\multicolumn{1}{l|}{$4.80(\pm 0.20)$}
   \\ \hline 
  HOT &
\multicolumn{1}{l|}{$5.17(\pm 0.21)$} &
\multicolumn{1}{l|}{$5.28(\pm 0.16)$} &
\multicolumn{1}{l|}{$5.17(\pm 0.18)$} &
\multicolumn{1}{l|}{$5.29(\pm 0.19)$} &
\multicolumn{1}{l|}{$5.32(\pm 0.18)$} &
\multicolumn{1}{l|}{$5.14(\pm 0.19)$} &
\multicolumn{1}{l|}{$5.00(\pm 0.19)$} &
\multicolumn{1}{l|}{\cellcolor{green!25}$5.45(\pm 0.18)$}
   \\ \hline 
  TM &
\multicolumn{1}{l|}{$4.32(\pm 0.22)$} &
\multicolumn{1}{l|}{$4.18(\pm 0.20)$} &
\multicolumn{1}{l|}{$4.29(\pm 0.21)$} &
\multicolumn{1}{l|}{$4.21(\pm 0.22)$} &
\multicolumn{1}{l|}{$4.62(\pm 0.20)$} &
\multicolumn{1}{l|}{\cellcolor{red!25}$3.89(\pm 0.22)$} &
\multicolumn{1}{l|}{$4.24(\pm 0.20)$} &
\multicolumn{1}{l|}{$4.26(\pm 0.22)$}
   \\ \hline
  WIA &
\multicolumn{1}{l|}{$5.25(\pm 0.22)$} &
\multicolumn{1}{l|}{$5.34(\pm 0.18)$} &
\multicolumn{1}{l|}{$5.22(\pm 0.21)$} &
\multicolumn{1}{l|}{$5.37(\pm 0.19)$} &
\multicolumn{1}{l|}{$5.54(\pm 0.19)$} &
\multicolumn{1}{l|}{$5.07(\pm 0.20)$} &
\multicolumn{1}{l|}{$5.13(\pm 0.20)$} &
\multicolumn{1}{l|}{$5.46(\pm 0.20)$}
   \\ \hline
  CCLAH &
\multicolumn{1}{l|}{$1.08(\pm 0.2)$} &
\multicolumn{1}{l|}{$0.77(\pm 0.22)$} &
\multicolumn{1}{l|}{$1.07(\pm 0.22)$} &
\multicolumn{1}{l|}{$0.78(\pm 0.20)$} &
\multicolumn{1}{l|}{$0.70(\pm 0.21)$} &
\multicolumn{1}{l|}{$1.14(\pm 0.20)$} &
\multicolumn{1}{l|}{$0.80(\pm 0.19)$} &
\multicolumn{1}{l|}{$1.04(\pm 0.23)$}
   \\ \hline
  CS &
\multicolumn{1}{l|}{$-0.04(\pm 0.16)$} &
\multicolumn{1}{l|}{$0.49(\pm 0.20)$} &
\multicolumn{1}{l|}{$0.06(\pm 0.17)$} &
\multicolumn{1}{l|}{$0.40(\pm 0.19)$} &
\multicolumn{1}{l|}{$0.41(\pm 0.17)$} &
\multicolumn{1}{l|}{$0.05(\pm 0.19)$} &
\multicolumn{1}{l|}{$0.21(\pm 0.18)$} &
\multicolumn{1}{l|}{$0.24(\pm 0.18)$}
   \\ \hline 
  CTA &
\multicolumn{1}{l|}{$3.86(\pm 0.19)$} &
\multicolumn{1}{l|}{$3.78(\pm 0.16)$} &
\multicolumn{1}{l|}{$3.78(\pm 0.18)$} &
\multicolumn{1}{l|}{$3.86(\pm 0.16)$} &
\multicolumn{1}{l|}{$3.80(\pm 0.17)$} &
\multicolumn{1}{l|}{$3.84(\pm 0.18)$} &
\multicolumn{1}{l|}{\cellcolor{green!25}$4.01(\pm 0.16)$} &
\multicolumn{1}{l|}{\cellcolor{red!25}$3.64(\pm 0.18)$}
   \\ \hline
\end{tabular}}
\caption{Summary of results for Study-1's classroom deployment. The short codes for outcome measures used here are defined in Table \ref{table:study1-measures}.}
\label{table:field_summary}
\end{table}

\subsubsection{Performance measures (RQ2)} For Homework Score, we see a contrast between \textit{G4: Solve-Then-Refine} (focusing on getting them to solve a problem first, then go to the LLM with questions), vs. G1, G2, G3 (which focused on using the LLMs while solving problems). The data suggest that G4-Solve-Then-Refine (encouraging self-first approach) may not have helped homework performance as much compared to the other forms of guidance (see Table \ref{table:field_summary}). 

\subsubsection{Perception of system} On average, we observe a small increase in "confidence in the ability of LLM to help" across different forms of guidance strategies and use approaches (RQ4). All conditions report moderately high ratings for "Helpfulness" for the given topic. However, they report it to be even higher on average for "Helpfulness for other topics", with the presence of G4-solve-then-refine standing out for being particularly helpful in general. The data suggest that the presence of G3-Metacognitive-Questioning reduced tolerance for mistakes and willingness to interact again.

\subsubsection{Perception of self} There is negligible change in the student's self-confidence (their confidence on the given topic) and this is consistent across different conditions. The students' confidence in their answers is similar across different conditions, however, G4-Solve-then-refine (or encouraging self-first approach) stands out - the absence of G4 in the overall guidance resulted in relatively higher confidence in their answers (RQ5). 

The findings from our formative deployment established the potential for differences between different guidance strategies (particularly between G4 and other strategies) and the resulting approach (self-first approach) to influence different measures related to performance and confidence. We further evaluate the effect of different interaction strategies in the following section. One key takeaway from the formative deployment was to introduce intermediate measures immediately after giving guidance to the learner.

\section{Study-2: Solving Math Problems in an Online Controlled Setting}
\label{section:mturk}

\begin{figure}
  \includegraphics[width=\textwidth]{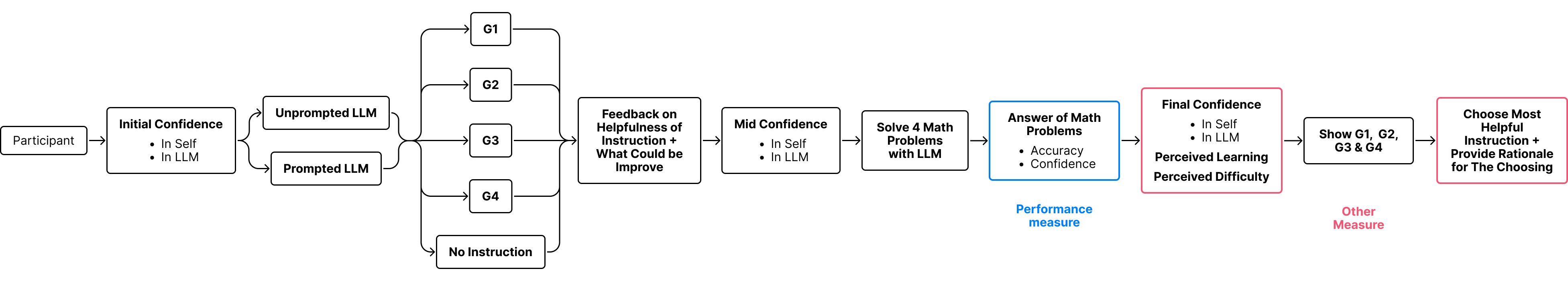}  
  \caption{Schematic of the experiment design for Study 2 with crowdworkers. 2 (\textit{LLM Chatbot}: Prompted vs. Unprompted) x 5 (\textit{Guidance strategy}: G1-List of Suggestions vs. G2-Example-based Instruction vs. G3-Metacognitive Questioning vs. G4-Solve then refine with LLM vs. No Instruction) between subjects for the type of chatbot and guidance received.} 
  \label{fig:prolific_design}
\end{figure}

To further explore the design space of the LLM-learner interaction, we designed a 2 (\textit{LLM Chatbot}: Prompted vs. Unprompted) x 5 (\textit{Guidance strategy}: G1-List of Suggestions vs. G2-Example-based Instruction vs. G3-Metacognitive Questioning vs. G4-Solve then refine with LLM vs. No Instruction) (Figure \ref{fig:prolific_design}) randomized controlled trial. Upon consenting to participate in the experiment, participants were asked to report their confidence in their ability to solve math problems, how frequently they used LLMs, and their confidence in using and trusting the responses generated by LLMs. The participants were then randomized to access a chatbot, either with or without a preprompt (detailed in Section \ref{section:prompt_polific}). They received one of four guidance strategies (explained in Section \ref{section:design}) or a basic description of how LLMs work (no instruction). After solving a common example problem with the LLM chatbot, they rated their confidence in using and trusting LLMs. Subsequently, they solved 4 math problems (randomly selected from a pool of 8 problems) with support from the LLM chatbot (Section \ref{section:prolific_task}). Finally, they reported their final confidence in their ability to solve math problems, their confidence in using and trusting LLM responses. The experiment flow just described is summarized and visualized in Figure \ref{fig:prolific_design}.

\subsection{Participants}
We recruited 356 participants, fluent in English, from Prolific \cite{palan2018prolific} \footnote{Prior research indicates that involving crowdworkers in user studies is effective \cite{kittur2008crowdsourcing}. It has been established that data obtained from crowdsourcing platforms can match the quality of traditional methods, including those involving undergraduate students and community samples \cite{follmer2017role}. Furthermore, crowdworkers have demonstrated their reliability as stand-ins for online learners in various contexts \cite{davis2018crowdworkers}.}. We did not have any special recruitment criteria other than fluency in English. Table \ref{table:participants} describes the participants.

\begin{table}
\centering

\begin{tabular}{|p{0.2\linewidth}||p{0.8\linewidth}|}
\hline
\textbf{Category} & \textbf{Results}\\
\hline\hline
Initial self-confidence in solving math problems & 1 (1.40\%), 2 (3.93\%), 3 (8.15\%), 4 (18.26\%), 5 (28.93\%), 6 (22.75\%), 7 (16.57\%)\newline \textit{where 1 is ``not confident at all'' and 7 is ``extremely confident''}\\
\hline
Prior use of LLMs &  Never (23.88\%), Once (20.51\%), Occasionally (34.55\%), Regularly (21.07\%)\\
\hline
Age  & 18-25 (43.26\%), 25-34 (41.57\%), 35-44 (10.96\%), 45-54 (2.81\%), 55-64 (1.12\%), 65+ (0.28\%)\\
\hline
Gender &  Male (47.75\%), Female (49.44\%), Non-binary / third gender  (1.97\%), Prefer not to say (0.84\%)\\
\hline
Education & Some high school or less (2.25\%), High School graduate (10.96\%), Some college credit with no degree (23.31\%), Associate degree (4.49\%), Bachelor's degree (41.85\%), Graduate or Professional degree (14.89\%), Prefer not to say (2.25\%)\\
\hline\hline
\multicolumn{2}{|c|}{\textbf{Total Participants = 356}} \\\hline
\end{tabular}
\caption{Demographic information of the participants recruited for Study-2 on Prolific.}
\label{table:participants}
\end{table}

\subsection{Task: Answering multiple choice math questions}
\label{section:prolific_task}
Each participant tackled two elementary-level math questions and two high-school-level math questions. The sequence of these tasks was randomized for each individual. From the MMLU benchmark dataset \cite{hendrycks2021ethics, hendryckstest2021}, we randomly chose 4 questions each from the elementary and high school math levels. The participants were then tasked with solving two questions (randomly sampled) from each level. 




\subsection{Steering LLM Responses with Pre-Prompts: \textit{Prompted} vs. \textit{Unprompted} LLM Chatbot}
\label{section:prompt_polific}

In the controlled setting, we also wanted to explore the differences between an LLM that was provided with vs without a system-prompt. The distinction is crucial for several reasons:

\begin{itemize}
    \item \textbf{Real-world relevance}: The unprompted version mirrors common LLM chatbots such as ChatGPT, representing the typical interaction users might have with publicly available LLMs.
    \item \textbf{Educational context}: A prompted LLM, on the other hand, could resemble a scenario in which instructors guide the model to provide specific types of response. This distinction was especially relevant as our classroom experiment (Section \ref{section:field}) could not employ unprompted LLMs due to ethical considerations.
\end{itemize}

Our design probe was a GPT-3-based chat interface (see Section \ref{section:llm_spec}), representative of other chat-based LLM tutors. Below, we detail the two distinct variants of LLM chatbots used in our study:

\subsubsection{GPT-3 with Pre-Prompt}
This variant was introduced to explore whether guiding the model with specific pre-prompts can offer advantages over generic out-of-the-box LLMs. We used the following pre-prompt for GPT-3:

\noindent
\begin{mdframed}[backgroundcolor=gray!20]
"You are a professional K12 math teacher helping students answer math questions.\\

\noindent
Give students explanations, examples, and analogies about the concept to help them understand. You should guide students in an open-ended way. Make the answer as precise and succinct as possible.\\

\noindent
You should help them in a way that helps them (1) learn the concept, (2) have confidence in their understanding, and (3) have confidence in your ability to help them. Before answering, reflect on how your answer will help you achieve goals (1), (2), and (3). Update your answer based on this reflection."
\end{mdframed}

\subsubsection{GPT-3 without Any Pre-Prompt}
This unprompted version served as a control, allowing us to contrast its performance and responses with the prompted version.


\subsection{Results}

\subsubsection{RQ2: Performance (Learner + AI performance in the task)} The performance (average score in the task) was not significantly different between different types of guidance ($\chi^2_{\text{Kruskal-Wallis}}(4) = 0.13$, $n = 356$, $p=1$) or whether the LLM was prompted or unprompted ($W_{\text{Mann-Whitney}} = 15025$, $n = 356$, $p=0.37$).

\begin{figure}
  \includegraphics[width=\textwidth]{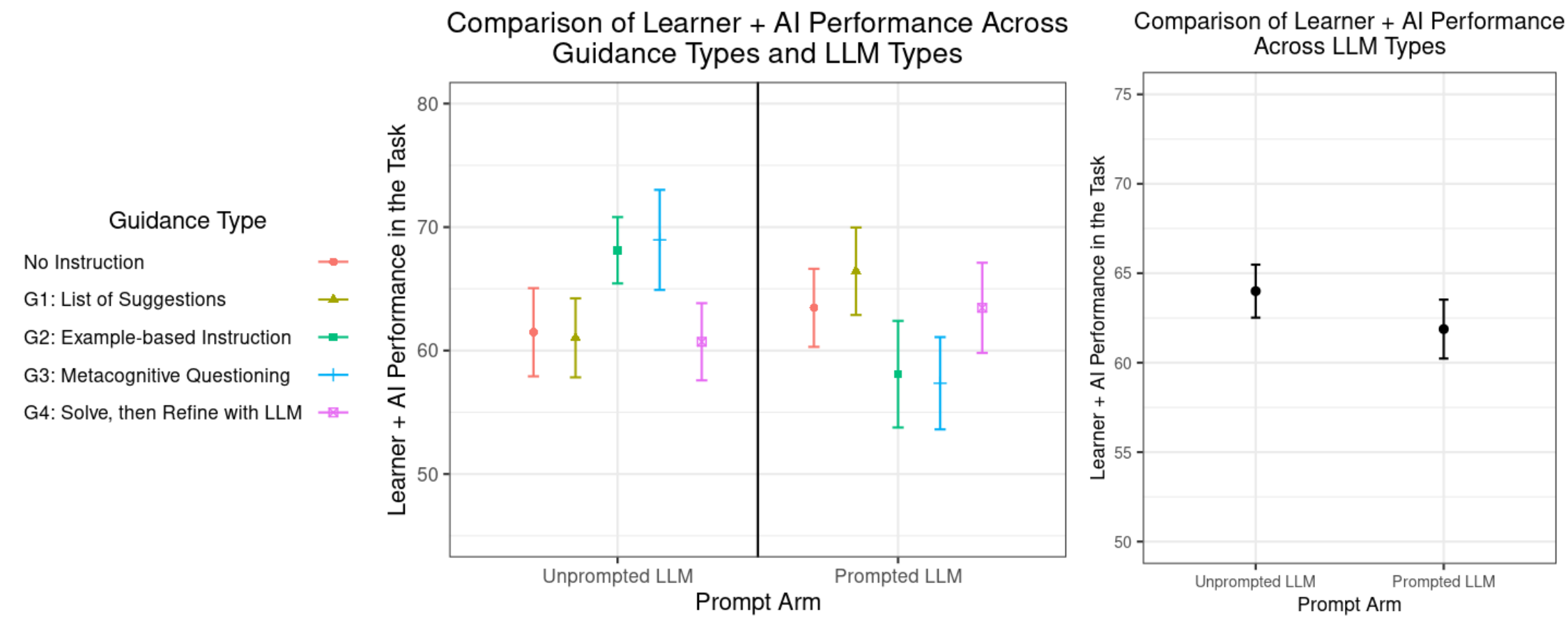}  
  \caption{Performance (operationalized as accuracy in the 4 questions) across the different conditions of the experiment.}  
  \label{fig:prolific_average}
\end{figure}

Exploratory analysis of subgroups (see Figure \ref{fig:prolific_average}) suggest that guidance strategies G2 (examples) and G3 (Metacognitive Questioning) appeared to improve performance when paired with unprompted LLM. This suggests that these guidance approaches may be effective in counterbalancing the absence of a pre-prompt given to the LLM. Specifically, the use of examples (G2) with unprompted LLM achieved an average score of \(68.13\% \pm 2.68\%\), compared to \(58.09\% \pm 4.32\%\) obtained with prompted LLM. A similar trend was observed for G3 (Metacognitive Questioning) across the two LLM conditions, calling for further investigation of the potential advantages of these guidance strategies when paired with an unprompted LLM.

\begin{figure}
  \includegraphics[width=\textwidth]{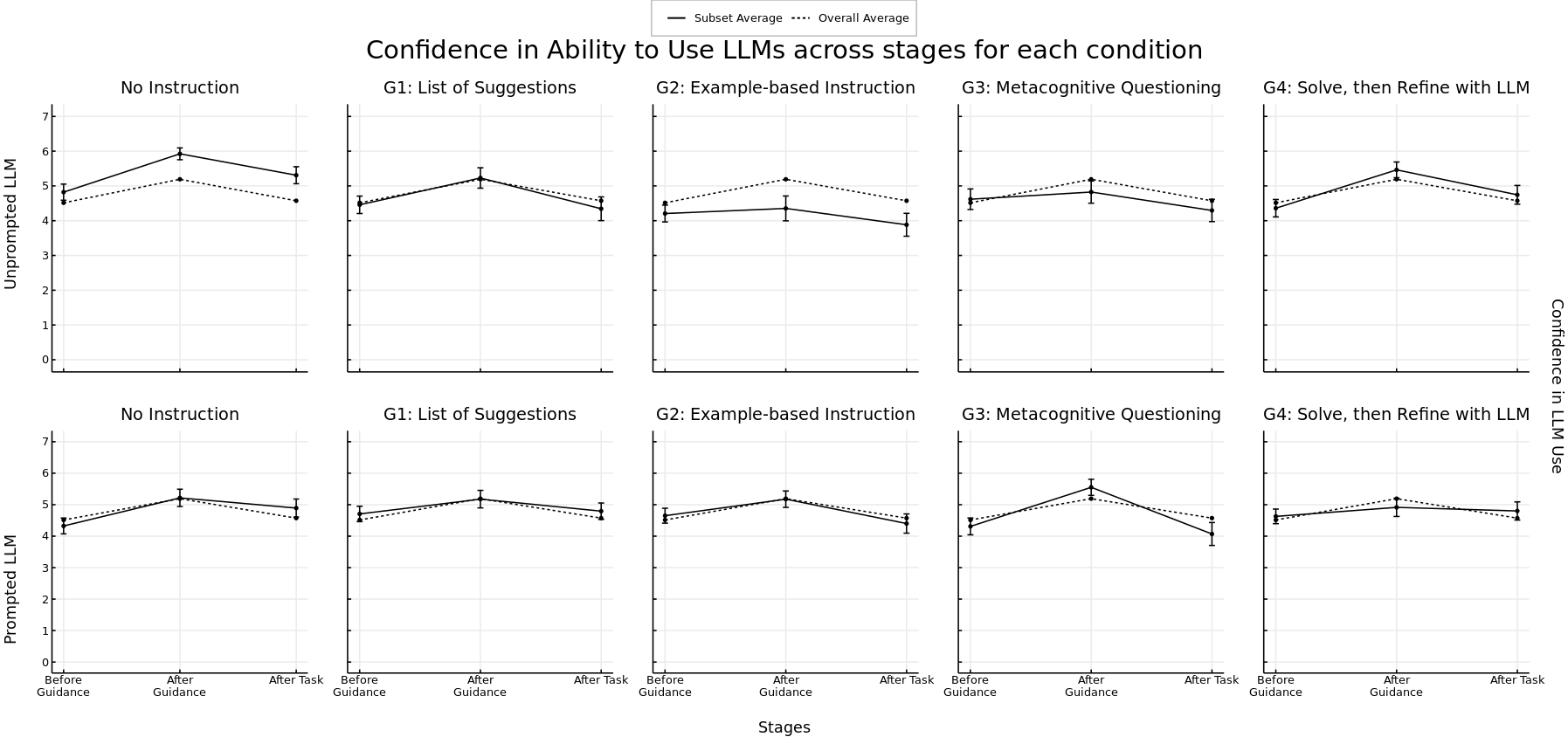}  
  \caption{This grid visualizes the participants' average confidence in their ability to use LLMs across different experimental conditions. Each cell represents a combination of LLM prompting (either "Prompted" or "Unprompted") and a specific guidance type. The gray dotted line in each cell represents the overall average confidence across all conditions, providing a point of reference. The overarching trend showcases varying levels of confidence shifts across different conditions.}  
  \label{fig:prolific_use_confidence}
\end{figure}

\subsubsection{RQ3: Confidence in ability to use LLMs to solve math problems over different stages of interaction (Figure \ref{fig:prolific_use_confidence})} The confidence is measured in three stages in the study in response to the prompt \textit{``I feel confident in my ability to use Large Language Models (LLMs) to assist me in solving math.''}. We measure this at the start of the study, immediately after receiving guidance and then after the problem-solving task. General confidence is moderately high, with the median being 4 out of 7. 

On average, across conditions, we observed significant differences in self-reported confidence in use of LLMs ($\chi^2_{\text{Friedman}}(2) = 72.82$, $n_{\text{pairs}} = 356$, $p < 0.001$), with growth ($p < 0.001$) between initial confidence ($\text{Mean} = 4.51$, $\text{Me} = 5$, $\text{SD} = 1.49$) and immediately after getting guidance ($\text{Mean} = 5.19$, $\text{Me} = 6$, $\text{SD} = 1.66$), and a significant drop ($p < 0.001$) after use ($\text{Mean} = 4.57$, $\text{Me} = 5$, $\text{SD} = 1.81$). The first part of this pattern suggests that people after receiving guidance, increase their confidence (as opposed to judging interaction as too complex), but after acquiring more experience in solving real tasks, their confidence slightly decreases. The reversion of the initial confidence level after the task is interesting, as it suggests that while guidance may temporarily boost confidence, the practical challenges of applying knowledge during problem-solving can temper this heightened confidence. Further exploratory analysis suggests that this pattern occurs for all the different forms of guidance, and even when there isn't guidance being provided, for the Prompted and the Unprompted LLM. 

Within the different conditions, these general trends are observed with varying degrees of intensity (see Figure \ref{fig:prolific_use_confidence}). For instance, the condition where the LLM is \textbf{Prompted} with \textbf{G4: Solve, then Refine with LLM} guidance shows the most pronounced retention of confidence after the problem-solving task. On the other hand, the \textbf{Unprompted} LLM condition with \textbf{G2: Example-based Instruction} shows the least initial boost in confidence post-guidance. These observations suggest potential nuanced interplays between LLM prompting and the guidance type in influencing learner confidence.

\begin{figure}
  \includegraphics[width=\textwidth]{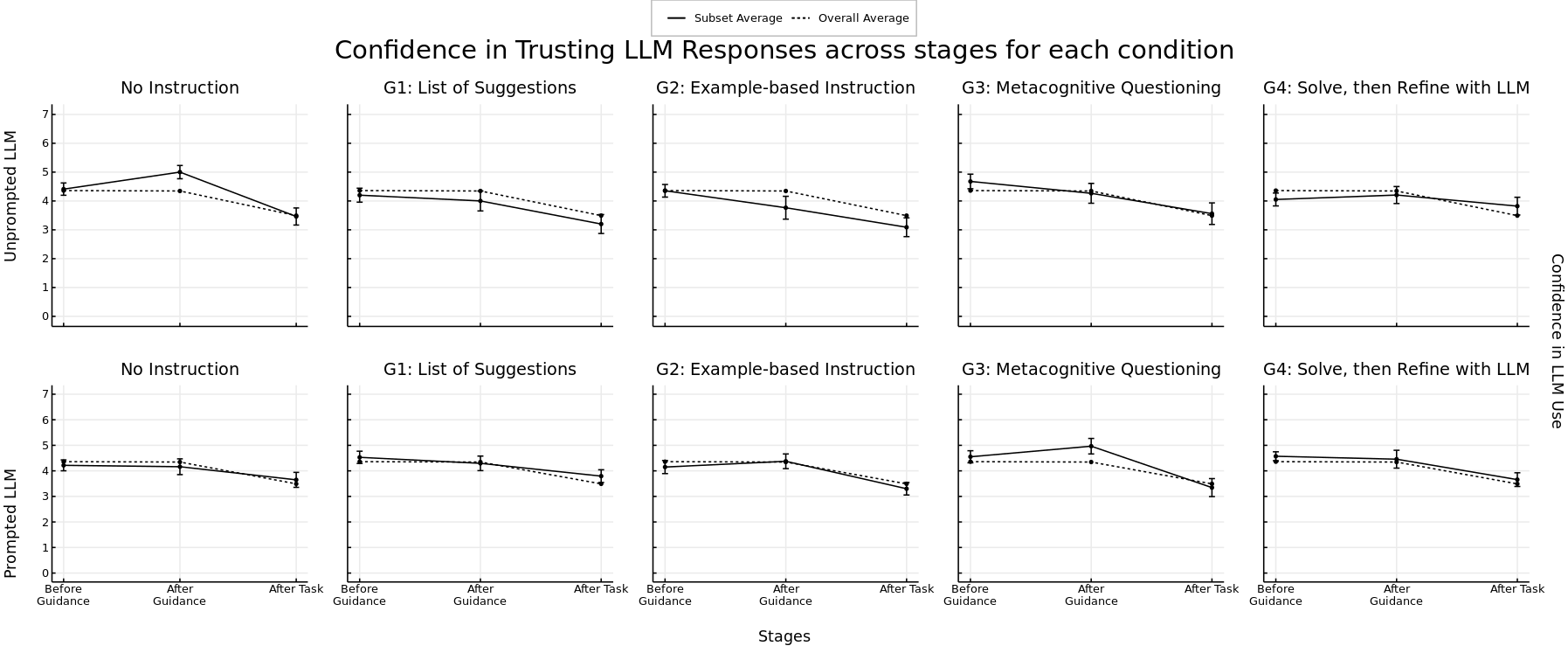}  
  \caption{This grid illustrates the average confidence of participants in trusting LLM responses across different experimental conditions. Each cell showcases a combination of LLM prompting and guidance type. The gray dotted line in each cell represents the overall average confidence across all conditions, providing a point of reference. The general trend indicates a decline in trust across most conditions, with a few exceptions.}  
  \label{fig:prolific_trust_llm}
\end{figure}

\subsubsection{RQ4: Trust in LLM's responses over different stages of interaction (Figure \ref{fig:prolific_trust_llm})}

On average, across conditions, we observed significant differences in reported trust in LLM responses ($\chi^2_{\text{Friedman}}(2) = 80.31$, $n_{\text{pairs}} = 356$, $p < 0.001$).

While the growth between initial trust ($\text{Mean} = 4.36$, $\text{Me} = 4$, $\text{SD} = 1.36$), immediately after getting guidance ($\text{Mean} = 4.35$, $\text{Me} = 5$, $\text{SD} = 1.89$) is not significant, there is a significant decline in final post-experience trust to LLM measure ($\text{Mean} = 3.49$, $\text{Me} = 3$, $\text{SD} = 1.80$) both compared to initial ($p<0.001$) and after instructions ($p<0.001$) measurements. This sharper decline might suggest that participants experience untrustworthy advice from LLM.

Further exploratory analysis within the different conditions generally confirms this observed trend (refer to Figure \ref{fig:prolific_trust_llm}). For example, in conditions where the LLM is \textbf{Unprompted} with \textbf{No Instruction} guidance, the decline in confidence after the task is the highest. However, there are exceptions. For instance, the \textbf{Unprompted} LLM condition with \textbf{G2: Example-based Instruction} shows the most significant decline in confidence immediately post-guidance, suggesting that certain guidance types might not instill trust in the LLM's responses.

\begin{figure}
  \includegraphics[width=\textwidth]{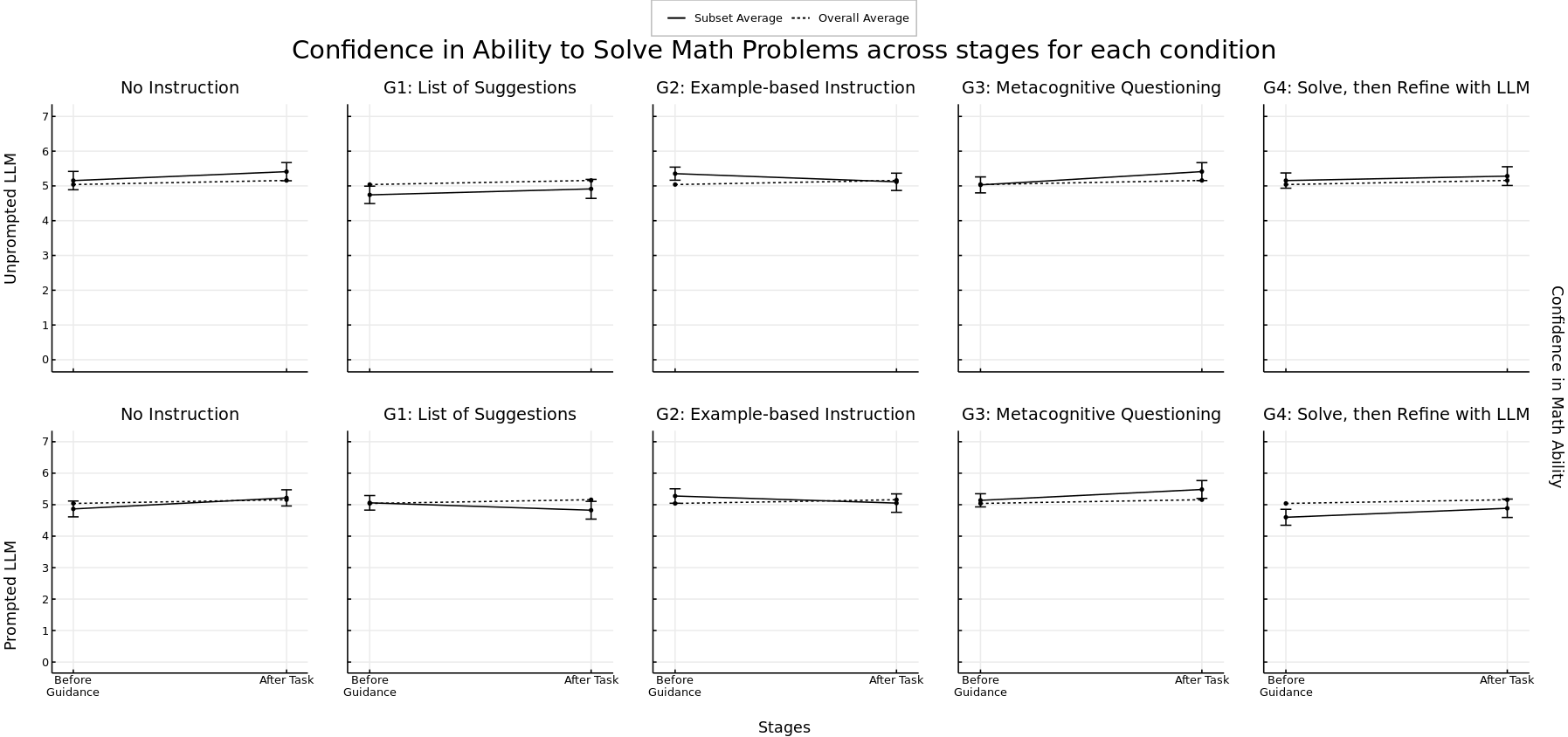}  
  \caption{The grid displays the average confidence of participants in their math-solving abilities across distinct experimental conditions. Each cell represents a combination of LLM prompting and a specific guidance type. The gray dotted line in each cell represents the overall average confidence across all conditions, providing a point of reference. The confidence levels across conditions appear to be stable with minor fluctuations.}  
  \label{fig:prolific_math_confidence}
\end{figure}

\subsubsection{RQ5: Changes in self-confidence in solving math problems before and after the problem-solving task with LLM (Figure \ref{fig:prolific_math_confidence})}

Our analysis reveals that participants' confidence in their math-solving abilities remains relatively stable throughout the interaction (see Figure \ref{fig:prolific_math_confidence}). The average confidence before receiving any guidance is $5.04$ ($\text{SD} = 1.41$). After they complete the problem-solving task, this confidence is $5.16$ ($\text{SD} = 1.62$). The observed difference between these two stages is not statistically significant with $p = 0.079$.

When exploring the individual experimental conditions, this stability in confidence is largely consistent. For instance, under the condition where the LLM is \textbf{Unprompted} and the guidance provided is \textbf{G3: Metacognitive Questioning}, we observe a slight increase in confidence. On the other hand, the \textbf{Unprompted} LLM condition combined with \textbf{G2: Example-based Instruction} guidance shows a minor decrease in post-task confidence. However, it's worth noting that these variations might not necessarily imply a significant impact on the participants' self-assessment of their math-solving abilities.

\section{Discussion}
\label{section:discussion}


Through the two studies, we investigated the effects of different guidance strategies on the learning process of students supported by an LLM tutor. In our discussion, we first summarize the key empirical findings and emphasize our contribution as it relates to design of Learner-LLM systems. We then provide some design considerations based on our findings, after which we conclude with limitations of our work and possible future work.

\subsection{RQ1: Impact of Guidance on Initiating Dialogue with LLM Tutor}
In our Formative Study, we observed distinct patterns in how students initiated dialogues with the LLM tutor, influenced by different types of guidance (G1--G4). Firstly, we noted a tendency for students to pose unrelated questions more frequently in the absence of \textit{G1-List of Suggestions}, \textit{G3-Metacognitive Questioning}, and \textit{G4-Solve then refine with LLM}. This could be interpreted in the light of existing research, which emphasizes the importance of structured guidance in collaborative environments \cite{lazakidou2010using, kim2018promoting, wu2021exploring}. The presence of structured guidance appears to channel student interactions more effectively towards relevant problem-solving. Conversely, the introduction of \textit{G2-Example based instruction} led to an increase in unrelated questions or attempts to break the chatbot. When students are presented with examples, they might test the system's capabilities beyond the task's scope, reflecting a curiosity-driven exploration \cite{gomez2021level, ten2021humans}.

Interestingly, G1, G2, and G4 were effective in reducing instances of students copy-pasting assignment questions directly, aligning with educational theories that suggest guided learning encourages deeper processing of information \cite{merrill1995tutoring, brydges2010comparing}. However, the introduction of \textit{G3-Metacognitive Questioning} showed an opposite effect, leading to more rephrasing of assignment questions. This aligns with education literature that highlights the role of metacognition in encouraging students to engage more deeply with the material, potentially prompting them to rephrase questions in their own words \cite{livingston2003metacognition, csen2009relationsip}. The guidance \textit{G4-Solve then refine with LLM} led to an increase in clarification questions as initial queries, consistent with the instruction given. This observation resonates with the principle of task-oriented communication, where specific guidance can direct collaborative efforts towards more productive and goal-oriented dialogue \cite{christie1952communication}.

\subsection{RQ2: Immediate Performance and Potential Long-term Learning with LLMs} Our formative study highlighted a potential design tension in LLM research: the balance between providing immediate assistance and ensuring long-term skill development for students. This fits into the long line of work studying the effects of automated hints on learner performance and learning \cite{marwan2019evaluation, corbett2001locus, fossati2015data, roy2016scale, rivers2017automated}. An interesting observation in the formative study was the contrast between G4: Solve-Then-Refine (focusing on getting them to solve a problem first, then go to the LLM with questions), vs G1, G2, G3 (which focused on using the LLMs while solving problems). There is suggestive evidence that the presence of G4-Solve-Then-Refine (encouraging self-first approach) may not have helped homework performance as much compared to the other forms of guidance (refer to Table \ref{table:field_summary}). However, the performance in the final exam without any LLM support (considered as a measure of learning) was similar between the different forms of guidance and learner approaches. A recent large-scale controlled study by Kumar \textit{et al.} \cite{kumar2023math} showed that when learners received explanations from GPT-4, the long-term benefits were greater for those who attempted problems on their own first before consulting LLM explanations. 

For Study-2, we did not find any significant differences between the guidance strategies or chatbot types on average, in terms of performance on the math problem-solving task with LLM (see Figure \ref{fig:prolific_average}). However, exploratory analysis looking at groups of Prompted vs. Unprompted LLM suggested that G2 (examples) and G3 (metacognitive questioning) improved performance when paired with unprompted LLM. This suggests that the right student guidance may offset the lack of a pre-prompt to the LLM. This finding is crucial when students independently use publicly available LLMs like ChatGPT. Instructors may guide this independent use by showing proper examples or encouraging deeper thinking about LLM use.

\subsection{RQ3 \& RQ4: Learner Confidence in Using LLMs and Trusting LLM Responses Across Different Stages of Problem-Solving Tasks}  
Formative study suggested a small increase in the confidence of the students in the ability of LLM to help, after solving assignment problems with LLM. However, there were no significant differences between the different guidance strategies and approaches of LLM use. In Study-2, when we looked at participants' confidence in using LLMs, we observed an interesting pattern. After receiving the guidance, the confidence of the people initially increased. However, as they gained more hands-on experience solving tasks with LLMs, their confidence dipped slightly. This poses questions for further research: Is this confidence drop a moment of realization about the challenges of seeking help, be it from LLMs or others? Might this realization push them to invest more effort in mastering the skill? Does the drop indicate a need for better guidance and support to ensure that users continue to take advantage of these tools for learning and assistance? Their trust in LLM's responses showed a similar trend. Although the growth between initial trust and immediately after receiving guidance is not significant, there is a significant decline in final post-task self-reported confidence in trusting LLM's responses. There could be complementary mechanisms behind this: People are either getting incorrect or misleading responses, or the responses are just not very helpful. This trend could also have been influenced by a mismatch in user expectations, often shaped by their familiarity with human-human interactions in problem-solving contexts. Learners might have project these expectations inaccurately when interacting with AI agents, leading to a potential disparity in the perceived effectiveness of the assistance. Existing research has highlighted that dynamics in human-agent conversations significantly differ from those in human-human interactions \cite{doyle2019mapping, clark2019makes}. This difference could account for the observed changes in confidence and trust levels, underscoring the need for aligning user expectations with the unique nature of AI-driven problem-solving support. Understanding these dynamics further could be pivotal in enhancing the effectiveness and acceptance of LLM-based learning tools. Further studies should investigate the qualitative aspects of learners' experiences to answer these questions.

\subsection{RQ5: Learners' Self-Assessment During the Problem-Solving Task} Across both studies, we did not find significant differences between the pre-task and post-task measures of learners' confidence in their ability to solve problems for the given topic with LLM assistance. Our findings contrast with some existing work in education that states mastery experiences (i.e., success in tasks) elevate self-confidence \cite{bandura1982self}. This could suggest that the perceived role of LLM in problem-solving might mediate the relationship between task success and self-confidence. This holds significance for the education research - mere problem solving with advanced tools does not translate to increased self-confidence. This underscores the need for complementary strategies in educational settings - such as reflection, discussion, or additional feedback mechanisms - to foster self-belief along with skill acquisition \cite{schunk1985self, margolis2005increasing, jackson2002enhancing}. In Study-2, participants reported moderately high confidence in their answers. But more important than just the average confidence is the calibration - whether their confidence is aligned with their actual accuracy. Approximately 23.53\% of participants in the \textit{Unprompted LLM} with \textit{G1: List of Suggestions} condition reported higher than median confidence in their answers, but achieved a lower than median score in the math problems. This misalignment echoes the broader educational concern of calibration in self-assessment - it is not just about how confident learners feel, but how that confidence matches up to their real-world performance \cite{marteau1989cognitive, gottlieb2022confidence}. For educators and technology developers, this underscores the importance of creating tools that not only assist in problem-solving but also foster accurate self-reflection and awareness in learners. Future research should delve deeper into strategies to better calibrate learners' self-perceptions when utilizing LLM-based tools.  

\subsection{Subjective Measures of Helpfulness, Tolerance for Mistakes, Willingness to Interact Again with the LLM Chatbot (Formative Study)}
Our formative study in the classroom allowed us to gather subjective ratings for different usability measures for LLM tutors. The presence of G3-Metacognitive-Questioning asked students to reflect on the questions they wanted to ask before attempting the problem, analogous to what they might as a teacher. One could hypothesize having clearer questions to ask could improve their interaction with the system as they have a clear sense of what to ask, to get useful information. However, the data suggested this might not be the case, or even the opposite might happen, as the presence of G3-Metacognitive-Questioning reduced tolerance for mistakes and willingness to interact again. This might happen by learners gaining unrealistic expectations about how LLMs could help, or by them having overly specific questions, whereas an LLM might be more beneficial for other reasons. For example, giving them space to self-explain or think through in a discussion might be helpful \cite{williams2016axis, williams2016revising}. All conditions report moderately high ratings for ``Helpfulness'' for the given topic. However, they report it to be even higher on average for ``Helpfulness for other topics'', with the presence of G4-solve-then-refine standing out as being particularly helpful in general.

\subsection{Methodological Design Choices for Educational LLM Interventions}

Each study helped to address specific aspects of the research questions. With Study-1, our objective was to explore how undergraduate students will interact with LLMs in a natural classroom setting, striving to achieve their course goals. However, this field setting constrained our design space, excluding more risky exploration strategies, potentially disadvantaging some students, and limiting the study sample size. We focused on rich qualitative feedback and interaction data, allowing a detailed examination of how interaction strategies shape student engagement with LLMs and treating quantitative assessments as formative. Consequently, Study-2, conducted with crowdworkers, aimed to build on the results of Study-1 and take a more direct approach to compare the guidance conditions. This allowed us to evaluate individual interaction strategies and the effect of an unprompted LLM, which was infeasible in Study-1 due to ethical reasons.

Although the classroom setting of Study-1 provided useful information on the real-world use of LLMs for learning, the inherent limitation of small sample sizes was a significant challenge \cite{emerson2012classroom}. To address this, we used a 2\(\times\)2\(\times\)2\(\times\)2 factorial experiment design, which assessed the presence or absence of four distinct guidance strategies, thus improving the statistical power \cite{dziak2012multilevel}.
This contrasted with the approach of Study-2, where the larger pool of participants allowed a broader comparison of strategies and the integration of subjective measures at various intervals \cite{kittur2008crowdsourcing}. These repeated measures could have led to high rates of attrition in the classroom setting. However, the validity of the results of Study-2 may not align with those of Study-1 \cite{kumar2023math}, where students interacted with LLMs as part of their coursework in a more authentic context.

\subsection{Future Work \& Limitations}
There is potential for further explorations of the design space of learner-LLM interactions. We could not experiment with pre-prompts in the classroom deployment due to ethical and sample size constraints, and the controlled online setting did not have a measure for long-term learning. Future studies could explore the effect of changing LLM pre-prompts on the longer-term learning outcomes of the students. Moreover, there is rich literature on how to guide people to use tools (especially in the context of learning) \cite{gersten1986direct, adams1996research, zendler2018effect, anderson2007supporting, van2010example, palincsar1986metacognitive, bourner2003assessing}. Future research can build on the guidance strategies proposed in this paper and explore the use of other designs to guide students, as well as to encourage different approaches. Furthermore, while designing learner-LLM interactions, it is crucial to ensure the right balance of learners' confidence in their answers while using LLMs, with actual accuracy of their answers. High confidence with low accuracy could signal over-reliance and overconfidence. Future research could look at designing interactions that ensure an optimal balance of confidence and accuracy.

There are limitations to the current work. Study 1 and 2 differed in various aspects, including study design, participant pool, and subject matter. Although the findings of both studies can be considered complementary, interpreting them together should be done with caution due to the distinct study settings. In our formative study, the constraints of small to medium-sized classroom field studies inherently limit the scope of broad statistical generalizations. Additionally, observations from our specific classroom setting of an introduction to database course may not be generalized to other classroom settings and audiences. For example, students with non-computer science backgrounds might interact with LLMs differently and differ in LLM-related trust and confidence. More studies are needed to see whether results from the online controlled setting generalize to learning tasks in other contexts and populations of online learners, who may differ from crowdworkers in their goals and motivation. A larger sample confirmatory study might help to reliably evaluate multiple interactions between student characteristics, experimental variables, achievement, and perception outcomes, which were described as exploratory in the current work. Additionally, our findings could benefit further from a rigorous qualitative analysis of conversations between learners and LLM chatbots, which can help explain the mechanisms behind some of the effects observed in our studies. Future educational interventions that use LLMs can adopt more adaptive experimental designs such as the Multiphase Optimization Strategy (MOST) or Sequential Multiple Assignment Randomized Trials (SMART) \cite{collins2007multiphase}. These designs offer the flexibility to make dynamic adjustments based on evolving data, which is crucial in educational settings where interaction effects and student responses can vary widely. Using such adaptive strategies and careful design decisions, researchers can better navigate the trade-offs between validity and statistical power, leading to more effective and contextually appropriate LLM interventions for learning.

Deploying guidance strategies for an LLM tutor carries several risks. Students may develop incorrect mental models of the LLM's capabilities \cite{kieras1984role}, particularly if the guidance (like G1-List of instructions or G2-Examples) does not accurately represent the operational boundaries of the LLM. The choice of examples in G2-Examples is critical; inappropriate examples might mislead students about the applicability or functionality of the LLM, potentially leading to frustration or inappropriate reliance on the LLM tutor \cite{kulesza2013too, passi2024appropriate}. In addition, introducing complex guidance strategies or excessive information at the beginning might overwhelm the students, leading to cognitive overload \cite{feldon2007cognitive}. This is particularly a concern with strategies requiring high initial engagement or pre-existing knowledge. The guidance strategies in this study did not specifically account for the diverse needs and backgrounds of all students, such as non-native English speakers \cite{seale2013learning, hultgren2019english}. Future work could explore ways to make our proposed guidance strategies personalized for students from varying backgrounds and more accessible to students with special needs.

\section{Conclusion}
\label{section:conclusion}

Through this work, we highlight complexities of designing learner-LLM interactions and underscore the critical role of instructors, teachers, and policymakers in guiding these interactions. We explore the design space of this interaction with \textit{four} guidance strategies resulting in \textit{two} different approaches of use, with a prompted vs. unprompted LLM, in a classroom and a controlled online setting. Even with our findings and rich data, we merely scratch the surface on the complexities of collaborative learning systems involving LLM agents. LLMs demonstrate remarkable flexibility in supporting various strategies, but also demand oversight to ensure the application of the most effective strategy in each unique learning scenario. Teachers are already playing a demanding role in these collaborative learning environments, ensuring proper use of LLMs. This highlights an emerging imperative for CSCW and HCI researchers: empirically studying diverse ways of guiding students in LLM use, thus empowering teachers relying largely on anecdotal evidence or grey literature for integrating LLMs in classrooms. 

Our research has shown the promise of simple yet effective interventions, such as including examples of LLM use in assignments, prompting students to reflect on their use of LLMs, and encouraging a `solve-first-then-consult' approach with LLMs. These insights are crucial for educators and educational technology developers alike, as they collaboratively shape the next generation of LLM tools to be intuitive and pedagogically effective for diverse learners. Ultimately, our work paves the way for more in-depth investigations and broadens the avenues for further research within the CSCW and HCI communities, emphasizing the dual task for researchers and designers in optimizing LLM implementation for genuine student understanding and engagement.
\begin{acks}
We thank the members of the Intelligent Adaptive Interventions Lab for their feedback on the project. We also thank Marko Choi and Hammad Sheikh for their help in developing the chat interface, which allowed us to collect data at scale. We acknowledge the financial support of Dr. Liut from the Learning \& Education Advancement Fund from the Office of the Vice-Provost, Innovations in Undergraduate Education, University of Toronto, Microsoft's Accelerating Foundation Model Research program, Natural Sciences and Engineering Research Council of Canada (NSERC) (\#RGPIN-2024-04348), as well as, Dr. William's  Natural Sciences and Engineering Research Council of Canada (NSERC) (\#RGPIN-2019-06968) and the Office of Naval Research (ONR) (\#N00014-21-1-2576) grants. Furthermore, we acknowledge an award from DARPA for AI Tools for Adult Learning awarded to TutorGen for QuickTA. 
\end{acks}

\bibliographystyle{ACM-Reference-Format}
\bibliography{ref}

\appendix

\section{Research Methods}

\subsection{Formative Study}

\subsubsection{LLM Model Specification}
\label{section:llm_spec}
\begin{itemize}
    \item \textbf{model version}: text-davinci-003
    \item \textbf{number of parameters}: 175B
    \item \textbf{date of use}: April 2023
\end{itemize}

Configuration Settings:
\begin{itemize}
    \item \textbf{temperature}: 0
    \item \textbf{max tokens}: 300
    \item \textbf{top-p}: 1
    \item \textbf{frequency penalty}: 0
    \item \textbf{presence penalty}: 0.6
\end{itemize}

Prompt Design:
\begin{quote}
   \textit{``The following is a conversation with a database instructor. The instructor helps the human solve assignment problems related to database. The instructor never explicitly gives the solution. The instructor also never writes the SQL query. The instructor would only provide brainstorms to possible solutions without providing any SQL statements. This rule should be enforced in the entirety of the conversation. Following are the ddl files to create table given as part of the assignment.''}
\end{quote}

Interaction Environment:
\begin{itemize}
    \item \textbf{environment}: A platform with a conversation interface hosted on university servers. The platform directly calls the OpenAI's API for generated contents.
\end{itemize}

\subsection{Online Controlled Study}

\subsubsection{LLM Model Specification}
\begin{itemize}
    \item \textbf{model version}: text-davinci-003
    \item \textbf{number of parameters}: 175B
    \item \textbf{date of use}: September 2023
\end{itemize}

Configuration Settings:
\begin{itemize}
    \item \textbf{temperature}: 0
    \item \textbf{max tokens}: 300
    \item \textbf{top-p}: 1
    \item \textbf{frequency penalty}: 0
    \item \textbf{presence penalty}: 0.6
\end{itemize}

Prompt Design:
\begin{itemize}
    \item \textbf{prompted LLM}:
\begin{quote}
   \textit{``You are a professional K12 math teacher helping students answer math questions.\\ \newline Give students explanations, examples, and analogies about the concept to help them understand. You should guide students in an open-ended way. Make the answer as precise and succinct as possible.\\ \newline You should help them in a way that helps them (1) learn the concept, (2) have confidence in their understanding, and (3) have confidence in your ability to help them. Before answering, reflect on how your answer will help you achieve goals (1) (2) (3). Update your answer based on this reflection.''}
\end{quote}
    \item \textbf{unprompted LLM}:
\begin{quote}
   \textit{``You are a helpful assistant.''}
\end{quote}
\end{itemize}

Interaction Environment:
\begin{itemize}
    \item \textbf{environment}: A platform with a conversation interface hosted by Microsoft Azure service. The platform directly calls the OpenAI's API for generated contents.
\end{itemize}

\end{document}